\documentclass[
prd,preprint,aps,endfloats,tightenlines,
showpacs,superscriptaddress,nofootinbib,
amssymb
]{revtex4}
\usepackage{graphicx}
\begin{document}
\setlength{\baselineskip}{16pt}
%
\title{
$I=2$ Two-Pion Wave Function and Scattering Phase Shift
}
%
\author{
Kiyoshi Sasaki
}
\altaffiliation[ Present address : ]{
Department of Physics, 
Tokyo Institute of Technology, Tokyo 152-8551, Japan
}
\affiliation{
Center for Computational Sciences,
University of Tsukuba, Tsukuba, Ibaraki 305-8577, Japan
}
\author{
Naruhito Ishizuka
}
\affiliation{
Center for Computational Sciences,
University of Tsukuba, Tsukuba, Ibaraki 305-8577, Japan
}
\affiliation{
Graduate School of Pure and Applied Sciences,
University of Tsukuba, Tsukuba, Ibaraki 305-8571, Japan
}
\date{ \today }
%
%
\begin{abstract}
We calculate a two-pion wave function
for the $I=2$ $S$-wave two-pion system
with a finite scattering momentum
and estimate the interaction range between two pions,
which allows us to examine the validity of a necessary condition
for the finite size formula
presented by Rummukainen and Gottlieb.
We work in the quenched approximation 
employing the plaquette gauge action for gluons
and the improved Wilson action for quarks
at $1/a=1.63\ {\rm GeV}$
on $32^3\times 120$ lattice.
The quark masses are chosen to give
$m_\pi = 0.420$, $0.488$ and $0.587 \ {\rm GeV}$.
We find that the energy dependence 
of the interaction range is small
and the necessary condition is satisfied 
for our range of the quark mass
and the scattering momentum, $k \le 0.16 \ {\rm GeV}$.
We also find that 
the scattering phase shift can be obtained 
with a smaller statistical error
from the two-pion wave function
than from the two-pion time correlator.
\end{abstract}
\maketitle
%
%
\section{ Introduction }
\label{sec:introduction}
Calculations of the scattering phase shift
represent an important step for expanding our understanding
of the strong interaction based on lattice QCD
to dynamical aspects of hadrons.
For the simplest case of the $I=2$ $S$-wave two-pion system,
several calculations have been reported
in Refs.~\cite{Aoki:2002ny,Kim:2004gh,RBC:2006kp,Li:2007ey,Yamazaki:2004qb,Beane:2005rj}.
These calculations employ the finite size formula
presented by L\"uscher~\cite{Luscher:1986pf:1991ux}
and the extension to the system with a non-zero total momentum
by Rummukainen and Gottlieb~\cite{Rummukainen:1995vs},
in which the scattering phase shift is related
to the energy eigenvalue on a finite volume.
In previous applications of the formula
the energy was calculated
from an asymptotic time behavior
of the two-pion time correlator.

The derivation of the finite size formula
assumes a condition $R < L/2$
for the two-pion interaction range $R$
and the lattice size $L$,
so that the boundary condition does not distort
the shape of the two-pion interaction.
It is important to examine the validity of this necessary condition
for reliable results of the scattering phase shift.
The CP-PACS collaboration
calculated the two-pion wave function
for the ground state of the $I=2$ $S$-wave two-pion system
and estimated the interaction range from it~\cite{Aoki:2005uf}.
In their case
the scattering momentum is highly small,
thus their work is an examination of the condition
for the scattering length.
They found that the interaction range is
$R\sim 1.3 - 1.6 \ {\rm fm}$
for the pion mass $m_\pi = 0.520 - 0.860\ {\rm GeV}$.
They also evaluated the scattering length 
and found it can be given 
with a smaller statistical error
from the wave function
than from the two-pion time correlator.

In the present work
we extend their work of the necessary condition
for the scattering length (the scattering momentum $k \sim 0$)
to that for the scattering phase shift ($k \ne 0$)
for the $I=2$ $S$-wave two-pion system.
In order to set a finite momentum for the ground state,
we consider a system
having the total momentum ${\bf P}=(2\pi/L){\bf e}_x$.
We calculate the two-pion wave function
and evaluate the scattering phase shift from it,
as carried out by the CP-PACS collaboration
for the scattering length.

This article is organized as follows.
In Sec.~\ref{sec:Finite size formula}
we briefly review the derivation
of the finite size formula
presented by Rummukainen and Gottlieb~\cite{Rummukainen:1995vs},
with emphasis on the role of the condition
for the interaction range.
The calculation method of the wave function 
and the simulation parameters
are given in Sec.~\ref{sec:Method of calculations}.
In Sec.~\ref{sub-sec:Wave functions} 
we present our results of the wave function
and estimate the interaction range.
The scattering phase shift from the wave function
is provided and compared
to those with the two-pion time correlator
in Sec.~\ref{sub-sec:Scattering phase from wave function}.
Our results of the scattering length
and the scattering phase shift at the physical quark mass 
are presented
in Sec.~\ref{sub-sec:Scattering length and phase shift at physical quark mass}.
Our conclusions are given 
in Sec.~\ref{sec:conclusion}.
Preliminary reports of the present work
were presented in Ref.~\cite{Sasaki:2007aa}.
The calculation was carried out on VPP5000/80
at the Academic Computing and Communications Center 
of University of Tsukuba.
%
%
\section{Finite size formula}
\label{sec:Finite size formula}
We briefly review the derivation of the finite size formula
presented by Rummukainen and Gottlieb~\cite{Rummukainen:1995vs},
with emphasis on the role of the condition
for the interaction range.
We restrict ourselves the formula for the $I=2$ $S$-wave two-pion system,
but the generalization can be easily carried out.
The formula has also been derived from another approaches
in Ref.~\cite{Kim:2005RG} and \cite{Chirist:2005RG}.
We follow, however,
the original derivation in Ref.~\cite{Rummukainen:1995vs}.

First we consider a wave function
in the infinite volume in the Minkowski space
defined by
\begin{equation}
  \Psi( x_1 ; x_2 )
  = \langle 0 | \pi^{+}(x_1)   \pi^{+}(x_2) 
              | \pi^{+}(p_1) , \pi^{+}(p_2) ; {\rm in} \rangle
\ ,
\label{eqn:wfM_Lab_org_def}
\end{equation}
in the space like region, $(x_1-x_2)^2 < 0$.
The state $|\pi^{+}(p_1), \pi^{+}(p_2); {\rm in} \rangle$
is an asymptotic two-pion state
with the four-dimensional momenta $p_1$ and $p_2$.
$\pi^{+}(x)$ is an interpolating operator for $\pi^+$
at the four-dimensional position $x = ( x^0, {\bf x} )$.
We assume that the two-pion interaction range $R$
is finite and the wave function satisfies
\begin{equation}
  ( \Box_j + m_\pi^2 ) \Psi( x_1 ; x_2 ) = 0
   \qquad \mbox{ for \ $j=1,2$ }
\ ,
\label{eqn:wfM_Lab_org_free}
\end{equation}
for $-(x_1-x_2)^2 > R^2$,
where $\Box_j$ is the d'Alembertian
with respect to the coordinate $x_j$.

In order to remove
a trivial dependence of the center of mass coordinate 
$X=(x_1+x_2)/2$,
we introduce a relative wave function defined by 
\begin{equation}
   \phi(x) =  \Psi( x_1 ; x_2 ) \cdot {\rm e}^{ i P \cdot X }
\ ,
\label{eqn:wfM_Lab_def}
\end{equation}
where $x = x_1 - x_2$ is the relative coordinate 
and $P = p_1 + p_2 = ( E, {\bf P} )$
is the total four-dimensional momentum.
From (\ref{eqn:wfM_Lab_org_free})
$\phi(x)$ satisfies
\begin{eqnarray}
  ( \Box - k^2 ) \phi(x) &=& 0 
\ ,  
\label{eqn:wfM_Lab_free.1} \\
  P \cdot \partial \phi(x) &=& 0
\ ,  
\label{eqn:wfM_Lab_free.2}
\end{eqnarray}
for $-x^2 > R^2$,
where $k^2$ is the scattering momentum defined by 
\begin{equation}
  k^2 = P^2 / 4 - m_\pi^2
      = ( E^2 - {\bf P}^2 )/4 - m_\pi^2
\ .
\label{eqn:k2_def}
\end{equation}
(\ref{eqn:wfM_Lab_free.1}) and (\ref{eqn:wfM_Lab_free.2}) 
also yield
\begin{equation}
  \Bigl[     \nabla^2  
          -  ( {\bf P} \cdot {\nabla} )^2 / E^2 
          +  k^2      
  \Bigr] \phi(x) = 0
\ .
\label{eqn:wfM_Lab_free.3}
\end{equation}

We can obtain a relation 
between the wave function and the scattering phase shift
by introducing a center of mass frame.
The wave function in the center of mass frame 
$\phi_{\rm CM}(x_{\rm CM})$ 
is related from that in the original frame $\phi(x)$
by the Lorentz transformation,
\begin{equation}
   \phi_{\rm CM}(x_{\rm CM}) = \phi(x)
\ .
\label{eqn:wfM_CM_def}
\end{equation}
$x_{\rm CM}$ is the coordinate
in the center of mass frame given by
\begin{eqnarray}
     x^0_{\rm CM} &=& \gamma ( x^0     - {\bf v}\cdot{\bf x} ) \ ,  \\
 {\bf x}_{\rm CM} &=& \hat{\gamma} [ {\bf x} - {\bf v} x^0 ]   \ ,
\label{eqn:Lorentz_boost_cord}
\end{eqnarray}
where ${\bf v}$ is the velocity ${\bf v} = {\bf P}/E$
and $\gamma$ is the Lorentz boost factor 
$\gamma = 1/\sqrt{ 1 - {\bf v}^2 } = E/\sqrt{P^2}$.
The operation $\hat{\gamma}[{\bf x}]$ is defined by
\begin{equation}
   \hat{\gamma}[ {\bf x} ]
          = \gamma {\bf x}_{ {}_\parallel }
                 + {\bf x}_{ {}_\perp     }
\ ,
\label{eqn:gamma_hat_def}
\end{equation}
where
${\bf x}_{{}_\parallel }$ and
${\bf x}_{{}_\perp     }$ are components of ${\bf x}$
parallel and perpendicular to the velocity ${\bf v}$, 
{\it i.e.},
${\bf x}_{{}_\parallel } = {\bf v} ({\bf x}\cdot{\bf v}) / {\bf v}^2$ and
${\bf x}_{{}_\perp     } = {\bf x} - {\bf x}_{{}_\parallel}$.

From (\ref{eqn:wfM_Lab_free.1}) and (\ref{eqn:wfM_Lab_free.2}),
$\phi_{\rm CM}(x)$ satisfies
\begin{eqnarray}
  ( \Box - k^2 ) \phi_{\rm CM}(x) &=& 0   
\ ,   
\label{eqn:wfM_CM_free.1} \\
  \frac{\partial}{ \partial x^0 } \phi_{\rm CM}(x) &=& 0 
\ ,
\label{eqn:wfM_CM_free.2}
\end{eqnarray}
for $-x^2 > R^2$.
Thus, $\phi_{\rm CM}(x^0,{\bf x})$
is independent of $x^0$
and 
\begin{equation}
    \phi         (                             0 ,              {\bf x}  )
  = \phi_{\rm CM}( - \gamma({\bf v}\cdot{\bf x}) , \hat{\gamma}[{\bf x}] )
  = \phi_{\rm CM}(                             0 , \hat{\gamma}[{\bf x}] )
  \qquad \mbox{ for \ $|{\bf x}| > R$ }
\ .
\label{eqn:wfM_CM_Lab_relation}
\end{equation}
In the following we always consider the wave function at $x^0=0$
and omit the argument for the relative time,
for example $\phi({\bf x})=\phi_{\rm CM}(\hat{\gamma}[{\bf x}])$
for (\ref{eqn:wfM_CM_Lab_relation}).
We also know that
the Helmholtz equation is satisfied : 
\begin{equation}
  ( \nabla^2 + k^2 ) \phi_{\rm CM}({\bf x}) = 0
  \qquad \mbox{ for \ $|{\bf x}| > R$ }
\ .
\label{eqn:wfM_CM_free.3}
\end{equation}

We can expand $\phi_{\rm CM}({\bf x})$
in terms of the spherical Bessel $j_{l}(x)$ 
and the Neumann function         $n_{l}(x)$ as
\begin{equation}
  \phi_{\rm CM}({\bf x})
     = \sum_{l=0}^{\infty} \sum_{m=-l}^{l}
         b_{lm}(k)
         \cdot
         Y_{lm}(\Omega)
         \Bigl(   \alpha_{l}(k) \cdot j_{l}(k|{\bf x}|) 
                +  \beta_{l}(k) \cdot n_{l}(k|{\bf x}|)   \Bigr)  
\ ,
\label{eqn:wfM_CM_exp}
\end{equation}
with some constant $b_{lm}(k)$,
where $\Omega$ is the spherical coordinate for ${\bf x}$.
The conventions of $j_{l}(x)$, $n_l(x)$ and $Y_{lm}(\Omega)$ 
agree with those of Ref.~\cite{MESSIAH:book}.
The coefficients $\alpha_{l}(k)$ and $\beta_{l}(k)$ 
yield the scattering phase shift
as $\tan\delta_{l}(k)=\beta_{l}(k)/\alpha_{l}(k)$.

Next we consider a wave function
on a periodic box $L^3$ in the Euclidian space ( on the lattice ),
which is defined by 
\begin{equation}
  \Psi^L ( {\bf x}_1, \tau ; 
           {\bf x}_2, \tau   )
  = \langle 0 | \pi^{+}({\bf x}_1,\tau)
                \pi^{+}({\bf x}_2,\tau) 
                      | \pi^{+} \pi^{+} ; E, {\bf P} \rangle
\ ,
\label{eqn:wfE_Lab_org_def}
\end{equation}
where $| \pi^{+} \pi^{+}; E, {\bf P} \rangle$
is an eigenstate with the total energy $E$ and momentum ${\bf P}$ 
on the lattice. 
The wave function is periodic with respect to
the position of either of the pions :
\begin{eqnarray}
     \Psi^{L}( {\bf x}_1 + {\bf n}_1 L, \tau ;
               {\bf x}_2 + {\bf n}_2 L, \tau   )
  =  \Psi^{L}( {\bf x}_1              , \tau ;
               {\bf x}_2              , \tau   )
\qquad \mbox{ for \ ${\bf n}_1$, ${\bf n}_2$ $\in \mathbb{Z}^3$ }
\ .
\label{eqn:wfE_Lab_org_BC}
\end{eqnarray}
We introduce a relative wave function
like as in the Minkowski space by
\begin{equation}
 \phi^{L}({\bf x}) 
    = \Psi^{L} ( {\bf x}_1, \tau ;
                 {\bf x}_2, \tau   ) 
      \cdot {\rm e}^{ E \tau - i {\bf P}\cdot{\bf X} }
\ ,
\label{eqn:wfE_Lab_def}
\end{equation}
with ${\bf x}=  {\bf x}_1 - {\bf x}_2    $
and  ${\bf X}=( {\bf x}_1 + {\bf x}_2 )/2$.
(\ref{eqn:wfE_Lab_org_BC}) and (\ref{eqn:wfE_Lab_def})
yield 
\begin{equation}
   \phi^{L}({\bf x})
       = (-1)^{  L/(2\pi) {\bf P}\cdot{\bf n}  }
              \cdot \phi^{L}( {\bf x} + {\bf n}L )
   \qquad \mbox{ for \ ${\bf n}$ $\in \mathbb{Z}^3$ }
\ .
\label{eqn:wfE_Lab_BC}
\end{equation}

In the derivation of the finite size formula,
it is assumed that the two-pion interaction range $R$
is smaller than one-half the lattice extent,
so that the boundary condition does not distort
the shape of the two-pion interaction.
With this assumption,
$\phi^{L}({\bf x})$ satisfies 
the same equation as for $\phi({\bf x})$ :
\begin{equation}
  \Bigl[     \nabla^2  
          -  ( {\bf P}\cdot{\nabla} )^2 / E^2 
          +  k^2      
  \Bigr] \phi^{L}({\bf x}) = 0
     \qquad \mbox{ for \ $R < |\hat{\gamma}[{\bf x}]| < L/2$ }
\ .
\label{eqn:wfE_Lab_free.1}
\end{equation}
The finite size formula is given
by solving (\ref{eqn:wfE_Lab_free.1})
under the boundary condition (\ref{eqn:wfE_Lab_BC}).
It is convenient 
to change the variable
from ${\bf x}$ to ${\bf y}=\hat{\gamma}[{\bf x}]$
and define a new function by 
\begin{equation}
     \phi^{L}_{\rm CM}({\bf y})
   = \phi^{L}         ({\bf x}) 
\ ,
\label{eqn:wfE_CM_def}
\end{equation}
to solve the differential equation (\ref{eqn:wfE_Lab_free.1}).
$\phi^{L}_{\rm CM}({\bf x})$ satisfies
\begin{eqnarray}
&&   
   \Bigl(  \nabla^2  +  k^2  \Bigr) \phi^{L}_{\rm CM}({\bf x}) = 0 
   \qquad \mbox{ for \ $R < |{\bf x}| < L/2$ }
\ ,
\label{eqn:wfE_CM_free.1} \\
&&
   \phi^{L}_{\rm CM}({\bf x})
      = (-1)^{  L/(2\pi) {\bf P}\cdot{\bf n}  }
          \cdot \phi^{L}_{\rm CM}( {\bf x} + \hat{\gamma}[{\bf n}]L )
    \qquad 
    \mbox{ for \ ${\bf n}$ $\in \mathbb{Z}^3$ }
\ .
\label{eqn:wfE_CM_BC}
\end{eqnarray}
Note that (\ref{eqn:wfE_CM_def})
is not the Lorentz transformation on the lattice, 
but it is merely a definition of the function 
with the change of the variable 
from ${\bf x}$ to ${\bf y}=\hat{\gamma}[{\bf x}]$
to solve (\ref{eqn:wfE_Lab_free.1}).

The general solution of (\ref{eqn:wfE_CM_free.1}) 
under (\ref{eqn:wfE_CM_BC})
can be written by 
\begin{equation}
  \phi^{L}_{\rm CM}({\bf x})
     =  \sum_{l=0}^{\infty}
        \sum_{m=-l}^{l}
               v_{lm}(k) \cdot G_{lm}({\bf x};k) 
\ , 
\label{eqn:wfE_solution}
\end{equation}
with some constant $v_{lm}(k)$.
$G_{lm}({\bf x};k)$ is given from the periodic Green function :
\begin{eqnarray}
&&  G({\bf x};k)
       =  \frac{ 1 }{ \gamma L^3 }
             \sum_{ {\bf q} \in \Gamma_{\bf P} }
                 \frac{1}{ {\bf q}^2 - k^2 }
                 {\rm e}^{ i {\bf q}\cdot{\bf x}  }
\ , 
\label{eqn:Green_def}
\\
&&
  \Gamma_{\bf P}
    = \biggl\{
        \ {\bf q} \ \biggl{|}
        \ {\bf q} = \hat{\gamma}^{-1}
                  \Bigl[ \frac{2\pi}{L} {\bf n} + {\bf P} / 2 \Bigr] \ ,
        \ \ {\bf n} \in \mathbb{Z}^3
      \ \biggr\}
\ ,
\label{eqn:def_Gamma}
\end{eqnarray}
as
\begin{equation}
  G_{lm}({\bf x};k) = {\cal Y}_{lm}(\nabla) G({\bf x};k)
\ ,
\label{eqn:Glm_def}
\end{equation}
where ${\cal Y}_{lm}({\bf x}) = |{\bf x}|^{l} \cdot Y_{lm}(\Omega)$
with the spherical coordinate $\Omega$ for ${\bf x}$.
The expansion of $G({\bf x};k)$
in terms of $j_{l}(x)$ and $n_{l}(x)$ is given by 
\begin{equation}
  G({\bf x};k)
   =    \frac{k}{4\pi} n_{0}(k|{\bf x}|)
      + \sum_{l=0}^{\infty} 
        \sum_{m=-l}^{l}
           g_{lm}(k;1)
           \cdot Y_{lm}(\Omega) j_{l}(k|{\bf x}|)
\ .
\label{eqn:G_Exp}
\end{equation}
The function $g_{lm}(k;1)$ is an analytic continuation of 
\begin{equation}
  g_{lm}(k;z) =
    \left( \frac{ i }{ k } \right)^l
    \frac{ 4\pi }{ \gamma L^3 }
       \sum_{ {\bf q} \in \Gamma_{\bf P} } 
          {\cal Y}_{lm}({\bf q})  
          \cdot ( {\bf q}^2 - k^2 )^{-z}
\ ,
\label{eqn:spherical_zeta_function}
\end{equation}
which is defined for ${\rm Re}(z) > (l+3)/2$, 
where the region $\Gamma_{\bf P}$ 
is defined by (\ref{eqn:def_Gamma}).
The explicit expansion for $G_{lm}({\bf x};k)$ 
with general $l$ and $m$ is not needed.
Note, however, that
$G_{lm}({\bf x};k)$ contains
$j_{l'}(k|{\bf x}|)$ for a whole range of $l'$ and 
$n_{l'}(k|{\bf x}|)$ with only $l'=l$
as known from (\ref{eqn:Glm_def}) and (\ref{eqn:G_Exp}).

The wave function in the Euclidian space
is related to that in the Minkowski space 
by the analytic continuation of the relative time $i x^0 \to x^4$. 
It seems that
$\phi^{L}({\bf x}) = \phi({\bf x})$ at $x^0=x^4=0$
and 
\begin{equation}
   \phi^{L}_{\rm CM}({\bf x}) = \phi_{\rm CM}({\bf x})
\ , 
\label{eqn:wfE_wfM_Lab_rel0}
\end{equation}
from (\ref{eqn:wfM_CM_Lab_relation}) and (\ref{eqn:wfE_CM_def}).
But this is not true,
because the degeneracies of the energy eigenstate
in the infinite and the finite volume are different.
(\ref{eqn:wfE_wfM_Lab_rel0}) should be changed to 
\begin{equation}
   \phi^{L}_{\rm CM}({\bf x}) 
      =  \sum_{l=0}^{\infty}
         \sum_{m=-l}^{l}
             C_{lm}(k) \cdot 
             Y_{lm}(\Omega)
             \phi^{(lm)}_{\rm CM}(|{\bf x}|)  
\ ,
\label{eqn:wfE_wfM_CM_rel}
\end{equation}
with some constant $C_{lm}(k)$,
where $\Omega$ is the spherical coordinate for ${\bf x}$.
$\phi_{\rm CM}^{(lm)}(|{\bf x}|)$ is the $lm$ component of 
$\phi_{\rm CM}({\bf x})$ defined by 
\begin{equation}
  \phi^{(lm)}_{\rm CM}(|{\bf x}|)
      =  \int {\rm d}\Omega\ 
              Y_{lm}^{*}(\Omega)
              \phi_{\rm CM}({\bf x}) 
\ .
\label{eqn:wfE_wfM_CM_rel2}  
\end{equation}
Substituting 
(\ref{eqn:wfM_CM_exp}) and (\ref{eqn:wfE_solution})
into (\ref{eqn:wfE_wfM_CM_rel}),
we obtain 
\begin{eqnarray} 
   \phi^{L}_{\rm CM}({\bf x})  
  &=& \sum_{l=0}^{\infty}
      \sum_{m=-l}^{l}  
           v_{lm}(k) \cdot G_{lm}({\bf x};k)  
\cr 
  &=& \sum_{l=0}^{\infty} 
      \sum_{m=-l}^{l} 
            D_{lm}(k) \cdot
            Y_{lm}(\Omega)
            \Bigl(    \alpha_{l}(k) \cdot j_l(k|{\bf x}|)  
                    +  \beta_{l}(k) \cdot n_l(k|{\bf x}|)   \Bigr)   
\ , 
\label{eqn:wfE_wfM_CM_rel3} 
\end{eqnarray} 
where $D_{lm}(k) = C_{lm}(k) b_{lm}(k)$.

In the present work
we consider only the two-pion state 
in the ${\bf A}_1^+$ representation
of the rotational group on the lattice,
which equals to $S$-wave 
up to angular momentum $l=2$.
If the scattering phase shift for $l \ge 2$ is very small
in the energy range under consideration,
$\beta_{l}(k)\sim 0$ for $l \ge 2$
in (\ref{eqn:wfE_wfM_CM_rel3}).
This also means that
$v_{lm}(k)\sim 0$ for $l \ge 2$,
because $G_{lm}({\bf x};k)$ contains $n_l(k|{\bf x}|)$
as discussed before.
This expectation is supported 
by our numerical simulation as shown later.

We use the expansion form of $G({\bf x};k)$ in (\ref{eqn:G_Exp})
to determine the allowed values of $k$, $D_{lm}(k)$ and $v_{00}(k)$ 
in (\ref{eqn:wfE_wfM_CM_rel3}).
Comparing 
the coefficients of $j_{0}(k|{\bf x}|)$ and $n_{0}(k|{\bf x}|)$
of both lines of (\ref{eqn:wfE_wfM_CM_rel3}),
we find
\begin{eqnarray}
   D_{00}(k) \cdot \alpha_0(k) 
        &=&  v_{00}(k) \cdot \frac{1}{\sqrt{4\pi}} g_{00}(k;1)
\ , 
\label{eqn:RG_formula_S1} \\
   D_{00}(k) \cdot \beta_0(k)  
        &=&  v_{00}(k) \cdot \frac{k}{4\pi}
\ ,
\label{eqn:RG_formula_S2}
\end{eqnarray} 
where $g_{00}(k;1)$ is defined from (\ref{eqn:spherical_zeta_function}).
Finally we obtain the finite size formula
by taking the ratio of 
(\ref{eqn:RG_formula_S1}) and (\ref{eqn:RG_formula_S2}) : 
\begin{equation}
  \frac{ \alpha_0(k) }{ \beta_0(k) }
  = \frac{1}{\tan \delta_0 (k)}
  = \frac{\sqrt{4\pi}}{k} g_{00}(k;1)
\ .
\label{eqn:RG-formula}
\end{equation}
The other components of (\ref{eqn:wfE_wfM_CM_rel3})
give only information for $D_{lm}(k)$. 
In the case of ${\bf P}={\bf 0}$,
this formula turns to be the formula presented 
by L\"uscher in Ref.~\cite{Luscher:1986pf:1991ux}.
%
%
\section{ Method of calculations }
\label{sec:Method of calculations}
%
%
\subsection{ Calculation of wave function }
\label{sub-sec:Calculation of wave function}
In the present work
we consider
the ground state of the $I=2$ $S$-wave two-pion system
with the total momentum
${\bf P}={\bf 0}$ and ${\bf P}=(2\pi/L){\bf e}_x$.
When the interaction between two pions is turned off, 
the scattering momentum $k$ defined by (\ref{eqn:k2_def}) takes
\begin{equation}
\begin{array}[b]{llll}
  k & = 0              &       & \mbox{ for \ ${\bf P}={\bf 0}$           } \ , \\
  k & = \pi/(\gamma L) & \quad & \mbox{ for \ ${\bf P}=(2\pi/L){\bf e}_x$ } \ . \\
\end{array}
\end{equation}
These values of $k$ are changed by the two-pion interaction.

In order to calculate the wave function
we construct the correlation function :
\begin{equation}
  F_{\pi\pi}( {\bf x}, \tau , \tau_s )
      = \langle 0 |      \Omega ( {\bf x}, \tau   )
                    \bar{\Omega}( {\bf P}, \tau_s )
        | 0 \rangle
\ .
\label{eqn:four-point-func}
\end{equation}
The operator
$\Omega({\bf x},\tau)$ is defined by
\begin{equation}
  \Omega({\bf x},\tau)
     = \sum_{\hat{R}}
       \frac{1}{L^3} \sum_{\bf X }
            {\rm e}^{ i {\bf P} \cdot {\bf X} }
                 \pi^{+} ( {\bf X} + \hat{R}[{\bf x}] , \tau )
                 \pi^{+} ( {\bf X}                    , \tau )
\ ,
\label{eqn:def_Omega}
\end{equation}
where $\pi^{+}({\bf x},\tau)$ is
an interpolating operator for $\pi^{+}$ at the position $({\bf x},\tau)$.
The operation $\hat{R}$ represents an element of
the cubic      group ($O_h   $) for ${\bf P}={\bf 0}$ and
the tetragonal group ($D_{4h}$) for ${\bf P}=(2\pi/L){\bf e}_x$.
The summation over $\hat{R}$
projects out ${\bf A}_1^+$ representation of these groups,
which equals to the $S$-wave state
up to the angular momentum 
$l=4$ for ${\bf P}={\bf 0}$ and 
$l=2$ for ${\bf P}=(2\pi/L){\bf e}_x$.

The operator $\overline{\Omega}({\bf P},\tau_s)$
in (\ref{eqn:four-point-func}) is defined by
\begin{equation}
  \overline{\Omega}({\bf P},\tau_s)
    = \frac{1}{N_R}
      \sum_{j=1}^{N_R}
             \left[  \pi^{+} ( {\bf P}, \tau_s ; {\xi }_j )
                     \pi^{+} ( {\bf 0}, \tau_s ; {\eta}_j ) 
             \right]^\dagger
\ ,
\label{eqn:def_Omega_source}
\end{equation}
where
\begin{equation}
 \pi^{+}( {\bf P}, \tau_s ; {\xi }_j )
    = \frac{1}{L^3}
      \Bigl[    \sum_{\bf x}
                    {\rm e}^{ i {\bf P}\cdot{\bf x} }
                    \bar{d}( {\bf x}, \tau_s )
                    \xi_j^{*} ({\bf x})        \Bigr]
      \gamma_5
      \Bigl[    \sum_{\bf y}
                    u( {\bf y}, \tau_s ) 
                    \xi_j   ({\bf y})          \Bigr]
\ .
\label{eqn:pi_xi}
\end{equation}
The operator  ${\pi}^{+}({\bf P}, \tau_s ; {\eta}_j  )$
is defined as ${\pi}^{+}({\bf P}, \tau_s ; {\xi }_j  )$
by changing $\xi_j({\bf x})$ to $\eta_j({\bf x})$.
The functions $\xi_j({\bf x})$ and $\eta_j({\bf x})$ are
$U(1)$ noise whose property is
\begin{equation}
  \lim_{ N_R \to \infty }
        \frac{1}{N_R} \sum_{j=1}^{N_R}
               \xi_j^{*} ({\bf x}) 
               \xi_j     ({\bf y})
  = \delta^3 ( {\bf x} - {\bf y} )
\ .
\end{equation}
In the present work we set $N_R=2$.

Neglecting the contributions from excited states,
we can extract the wave function by
\begin{equation}
  \phi^{L}({\bf x})
    = \frac{ F_{\pi\pi}( {\bf x}   , \tau_0 , \tau_s ) }
           { F_{\pi\pi}( {\bf x}_0 , \tau_0 , \tau_s ) }
\ ,
\label{eqn:extr_wave_function}
\end{equation}
in the large $\tau_0$ region,
introducing a reference position ${\bf x}_0$.
We note that 
$\phi^{L}({\bf x})$ at all positions ${\bf x}$
are not independent.
The number of independent positions is
$(L+2)(L+4)(L+6)/48$ for ${\bf P}={\bf 0}$ and
$(L+2)^2   (L+4)/16$ for ${\bf P}=(2\pi/L){\bf e}_x$,
owing to the boundary condition (\ref{eqn:wfE_Lab_BC})
and the rotational symmetry on the lattice,
$\phi^{L}({\bf x}) = \phi^{L}(\hat{R}[{\bf x}])$.

In the present work 
we attempt to extract the energy of the two-pion system
and the scattering phase shift from the wave function.
For a comparison, 
we also evaluate them 
from the two-pion time correlator :
\begin{equation}
  G_{\pi\pi}( \tau , \tau_s )
    = \frac{1}{L^3} \sum_{\bf x} 
        F_{\pi\pi}( {\bf x} , \tau , \tau_s )
\ , 
\label{eqn:def_pp_tfunc} 
\end{equation}
as done in the previous works of the scattering phase shift.
We also calculate
the time correlator
for the pion and the $\rho$ meson,
\begin{eqnarray}
  G_{\pi}( \tau , \tau_s ) 
  &=& \frac{1}{N_R} \sum_{j=1}^{N_R} 
      \frac{1}{L^3} \sum_{\bf x}
           \langle 0 | \pi^+ ( {\bf x}, \tau )  
                \Bigl( \pi^+ ( {\bf 0}, \tau_s ; \xi_j ) \Bigr)^\dagger 
           | 0 \rangle 
\ ,
\\
  G_{\rho}( \tau , \tau_s ) 
  &=& \frac{1}{N_R} \sum_{j=1}^{N_R} 
      \frac{1}{L^3} \sum_{\bf x}
      \sum_{a=1,2,3}
           \langle 0 | \rho_a ( {\bf x}, \tau )  
                \Bigl( \rho_a ( {\bf 0}, \tau_s ; \xi_j ) \Bigr)^\dagger 
           | 0 \rangle 
\ ,
\label{eqn:tw-point-pi_rho}
\end{eqnarray}
which are used to extract the masses.
In (\ref{eqn:tw-point-pi_rho})
$\rho_a ({\bf x},\tau)$ is an interpolating operator for the $\rho$ meson 
with the polarization $a$ at the position $({\bf x},\tau)$ 
and $\rho_a ( {\bf 0}, \tau_s ; \xi_j)$ is given 
by substituting $\gamma_a$ for $\gamma_5$ in (\ref{eqn:pi_xi}).
%
%
\subsection{ Simulation parameters }
\label{sub-sec:Simulation parameters}
Our simulation is carried out in the quenched approximation
with the plaquette gauge action at $\beta=5.9$.
Gauge configurations are generated with
the $5$-hit pseudo-heat-bath algorithm
and the over-relaxation algorithm mixed in the ratio of $1:4$.
This combination is called
a sweep and the physical quantities are measured every $200$ sweeps
after $2000$ sweeps for the thermalization.

We use the improved Wilson action
for quarks~\cite{Sheikholeslami:1985ij}.
The clover coefficient $C_{SW}$ is chosen
to the mean-field improved value defined by
\begin{equation}
  C_{SW}
  = \langle U_\Box \rangle^{-3/4}
  = ( 1 - 2/\beta )^{-3/4}
  = 1.364
\ , 
\end{equation}
where $\langle U_\Box \rangle$ 
is the $1\times 1$ Wilson loop 
which is evaluated 
in one-loop perturbation theory.
The quark propagators are calculated
with the Dirichlet boundary condition imposed in the time direction
and the periodic boundary condition in the spatial one.
The source operator $\overline{\Omega}({\bf P},\tau_s)$
in (\ref{eqn:four-point-func})
is set at $\tau_s=20$ 
to avoid effects from the time boundary.

The lattice cutoff is estimated 
as $1/a = 1.631(45)\ {\rm GeV}$ ($a=0.1208(33)\ {\rm fm}$)
from $m_\rho$.
The lattice size is $32^3 \times 120$,
which corresponds to the spatial extent $3.87\ {\rm fm}$
in the physical unit.
We choose three quark masses to give
$m_\pi=0.420$, $0.488$ and $0.587 \ {\rm GeV}$ .
The numbers of configurations are $400$, $212$ and $212$
for each quark mass,
which are generated independently.
The masses of the pion and $\rho$ meson calculated
from the time correlator 
are listed in Table~\ref{tbl:mass-param}.
%
%
\section{ Results }
\label{sec:Results}
%
%
\subsection{ Wave functions }
\label{sub-sec:Wave functions}
\begin{figure}[h]
\begin{tabular}{ll}
\multicolumn{2}{l}{ (a) ${\bf P}={\bf 0}$ } \\
\includegraphics[width=85mm]{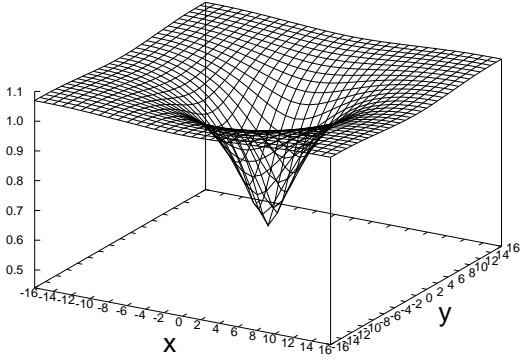}
\includegraphics[width=85mm]{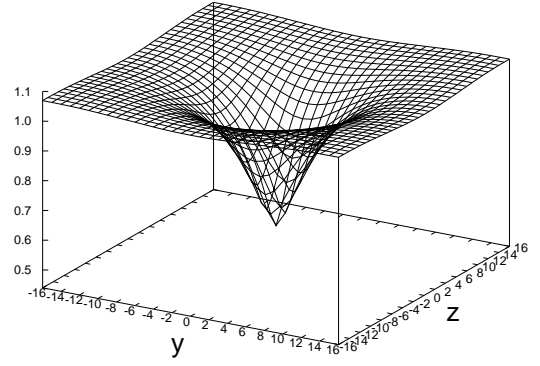} &
\\ \\
\multicolumn{2}{l}{ (b) ${\bf P}=(2\pi/L){\bf e}_x$ }  \\
\includegraphics[width=85mm]{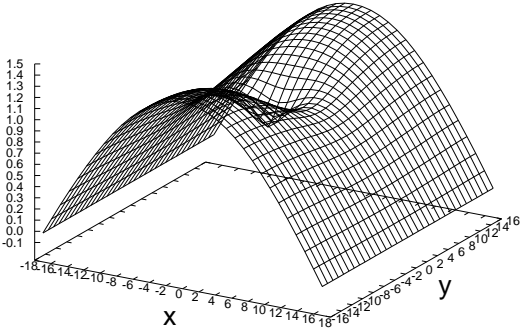}
\includegraphics[width=85mm]{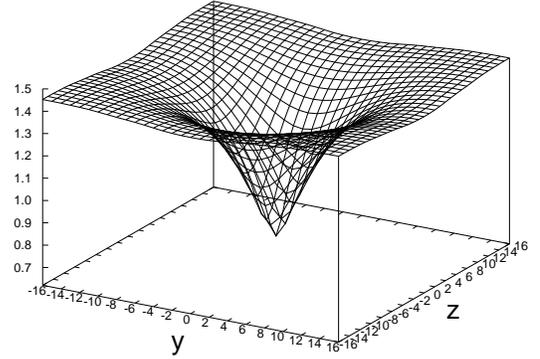} &
\end{tabular}
\caption{
Two-pion wave functions $\phi^{L}_{\rm CM}({\bf x})$
at $m_\pi = 0.420\ {\rm GeV}$
for the total momentum
${\bf P}={\bf 0}$ and ${\bf P}=(2\pi/L){\bf e}_x$.
The left and right panels for each momentum show
$\phi^{L}_{\rm CM}({\bf x})$
on $xy$-plane at $z=0$ and $yz$-plane at $x=0$.
}
\label{fig:wave-function-3d}
\newpage
\end{figure}

We obtain the wave function $\phi^{L}({\bf x})$
by (\ref{eqn:extr_wave_function}) 
and transform it to
$\phi^{L}_{\rm CM}({\bf x})=\phi^{L}(\hat{\gamma}^{-1}[{\bf x}])$.
In the transformation
we calculate the Lorentz boost factor by $\gamma=\sqrt{E^2-{\bf P}^2}$
with the energy $E$ extracted from the two-pion time correlator.
In Fig.~\ref{fig:wave-function-3d} 
we show $\phi^{L}_{\rm CM}({\bf x})$
at $m_\pi=0.420\ {\rm GeV}$
for the total momentum
${\bf P}={\bf 0}$ and ${\bf P}=(2\pi/L){\bf e}_x$.
Here we set the reference position ${\bf x}_0=(7,5,2)$
and $\tau_0 - \tau_s = 40$ in (\ref{eqn:extr_wave_function}),
confirming that the wave function
does not depend on the choice of
$\tau_0$ for $\tau_0 - \tau_s > 32$.
In the figure
the left and right panels for each momentum show
the wave functions on $xy$-plane at $z=0$ and $yz$-plane at $x=0$.
While the boundary conditions for ${\bf P}={\bf 0}$ 
are the periodic conditions in all directions,
those for ${\bf P}=(2\pi/L){\bf e}_x$
are anti-periodic in the $x$ direction 
and periodic in the other directions 
as known from (\ref{eqn:wfE_CM_BC}).
These features are clearly shown in the figure.

\begin{figure}[h]
\begin{tabular}{ll}
\multicolumn{2}{l}{ (a) ${\bf P}={\bf 0}$ } \\
\includegraphics[width=85mm]{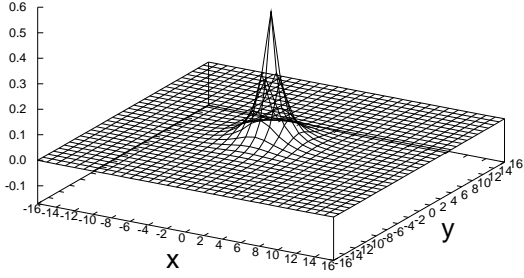} &
\includegraphics[width=85mm]{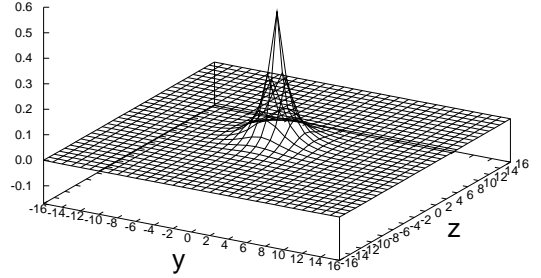}
\\ \\
\multicolumn{2}{l}{ (b) ${\bf P}=(2\pi/L){\bf e}_x$ } \\ 
\includegraphics[width=85mm]{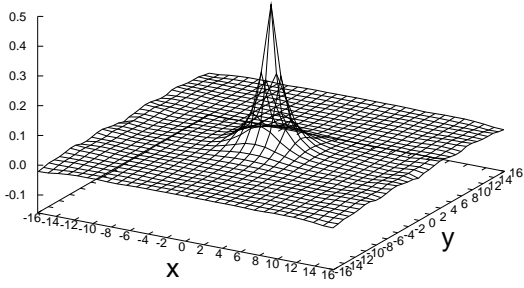} &
\includegraphics[width=85mm]{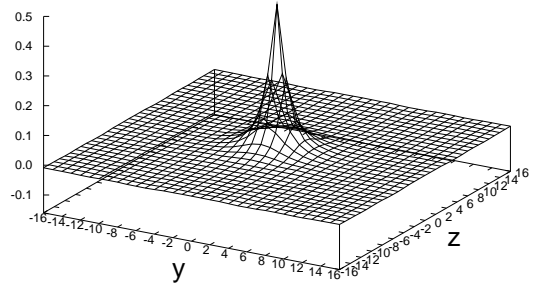}
\end{tabular}
\caption{
$V^{L}_{\rm CM}({\bf x})$ 
at $m_\pi = 0.420\ {\rm GeV}$
for the total momentum
${\bf P}={\bf 0}$ and ${\bf P}=(2\pi/L){\bf e}_x$.
The left and right panels for each momentum show
$V^{L}_{\rm CM}({\bf x})$ 
on $xy$-plane at $z=0$ and $yz$-plane at $x=0$.
}
\label{fig:Vx-3d}
\newpage
\end{figure}

We consider the two-pion interaction from the ratio :
\begin{equation}
  V^{L}_{\rm CM}({\bf y}) 
      = \frac{ \nabla^2_y  \phi^{L}_{\rm CM}({\bf y}) }
             {             \phi^{L}_{\rm CM}({\bf y}) }
      = \frac{ 1 }{ \phi^{L}({\bf x}) }
        \Bigl(  \nabla^2  
                   -  ( {\bf P}\cdot{\nabla} )^2 / E^2 
        \Bigr) \phi^{L}({\bf x})
\ ,
\label{eqn:V_def}
\end{equation}
where ${\bf y}=\hat{\gamma}[{\bf x}]$
and $\nabla^2_y$ is the Laplacian with respect to ${\bf y}$.
Away from the two-pion interaction range, 
we expect that 
$V^{L}_{\rm CM}({\bf x})$
is independent of ${\bf x}$ and equals to $-k^2$. 
In the calculation of the ratio,
we rewrite (\ref{eqn:V_def}) by
\begin{equation}
  V^{L}_{\rm CM}({\bf y})
    = \frac{1}{ \phi^{L}({\bf x}) }
      \left(
             \frac{1}{\gamma^2} \frac{\partial^2}{\partial x_1^2} 
           +                    \frac{\partial^2}{\partial x_2^2} 
           +                    \frac{\partial^2}{\partial x_3^2} 
      \right) \phi^{L}({\bf x})
\ ,
\label{eqn:V_rewrite}
\end{equation}
and adopt the naive numerical derivative,
\begin{equation}
  \frac{\partial^2}{\partial x_i ^2} f({\bf x})
   =
          f( {\bf x} + \hat{i} ) 
      +   f( {\bf x} - \hat{i} )
      - 2 f( {\bf x}           )
\ ,
\end{equation}
where the Lorentz boost factor 
$\gamma$ is calculated with $E$
extracted from the two-pion time correlator.
In Fig.\ref{fig:Vx-3d}
$V^{L}_{\rm CM}({\bf x})$ is plotted for the same parameters 
as for Fig.~\ref{fig:wave-function-3d}.
We find that
the ratio is almost constant for $|{\bf x}| > 10$
both for
${\bf P}={\bf 0}$ and 
${\bf P}=(2\pi/L){\bf e}_x$.
We also observe 
a strong repulsive interaction near the origin
consistent with 
the negative scattering phase of the $I=2$ two-pion system.

Here, we note a physical meaning of 
$V^{L}_{\rm CM}({\bf x})$ in (\ref{eqn:V_def}).
It seems that
$V^{L}_{\rm CM}({\bf x})$
equals to the corresponding ratio in the Minkowski space
defined by 
\begin{equation}
  V_{\rm CM}({\bf x}) 
     = \frac{ \nabla^2 \phi_{\rm CM}({\bf x}) }
            {          \phi_{\rm CM}({\bf x}) }
\ , 
\label{eqn:V_CM_M_def}
\end{equation}
which approximately takes the potential 
of the two-pion interaction in the non-relativistic limit.
This, however, is not true 
for ${\bf P}\not={\bf 0}$
in the interaction region $|{\bf x}|<R$.
Ignoring the difference of the degeneracy 
of the energy eigenstate
in the infinite and the finite volume,
the wave function in the Euclidian space, 
which is calculated in the present work,
is related to that in the Minkowski space
by the analytic continuation
as $\phi^{L}({\bf x}) = \phi({\bf x})$ at the relative time $x^0=0$.
But a relation between 
$\phi^{L}_{\rm CM}({\bf x})$ 
and $\phi_{\rm CM}({\bf x})$ 
is 
\begin{equation}
    \phi^{L}_{\rm CM}(\hat{\gamma}[{\bf x}])
  = \phi^{L}({\bf x})
  = \phi({\bf x}) 
  = \phi_{\rm CM}( - \gamma({\bf v}\cdot{\bf x}), \hat\gamma[{\bf x}])
\ ,
\end{equation}
which is not a relation at $x^0=0$.
Thus, there is no simple relation between
$V^{L}_{\rm CM}({\bf x})$ in (\ref{eqn:V_def}) and 
$V_{\rm CM}({\bf x})$     in (\ref{eqn:V_CM_M_def})
for ${\bf P}\not={\bf 0}$.
Only for the region $|{\bf x}|>R$,
we give a simple relation :
$V^{L}_{\rm CM}({\bf x}) = V_{\rm CM}({\bf x}) = - k^2$,
due to the $x^0$-independence of the wave function,
$\phi_{\rm CM}(x^0,{\bf x})=\phi_{\rm CM}(0,{\bf x})$.

We now consider the two-pion interaction range $R$.
In the quantum field theory
the wave function does not strictly satisfy 
the Helmholtz equation (\ref{eqn:wfE_CM_free.1}),
even for the large $|{\bf x}|$ region.
Hence,
with $k$ obtained from the two-pion time correlator,
$V^{L}_{\rm CM}({\bf x}) + k^2$ 
shows a small tail at large $|{\bf x}|$.
We may take the wave function
as satisfying the Helmholtz equation, 
if $V^{L}_{\rm CM}({\bf x}) + k^2$ 
is sufficiently small compared with $k^2$. 
In the present work
we take an operational definition
of the range $R$ as the scale,
where
\begin{equation}
  U^{L}_{\rm CM}({\bf x})
      = \frac{ V^{L}_{\rm CM}({\bf x}) + k^2 }{ k^2 }
      = \frac{ ( \nabla^2 + k^2 ) \phi^{L}_{\rm CM}({\bf x}) }
             {              k^2   \phi^{L}_{\rm CM}({\bf x}) }
\ ,
\label{eqn:U_def}
\end{equation}
is small enough compared with the statistical error.
With this definition
we expect that
the systematic error of the scattering phase shift 
from the interaction tail
is smaller than the statistical error of the scattering phase shift.

\begin{figure}[h]
\begin{tabular}{cc}
(a) ${\bf P}={\bf 0}$   &
(b) ${\bf P}=(2\pi/L){\bf e}_x$  \\
\includegraphics[width=80mm]{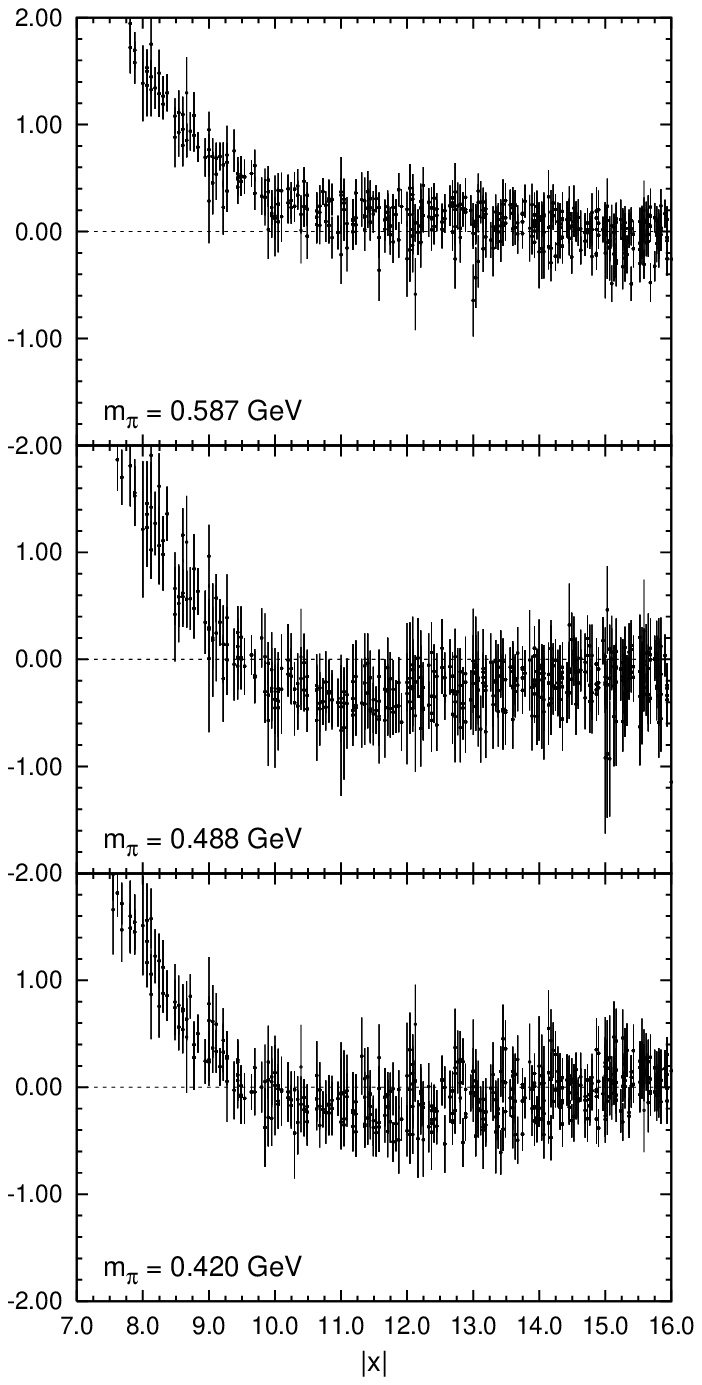}   &
\includegraphics[width=80mm]{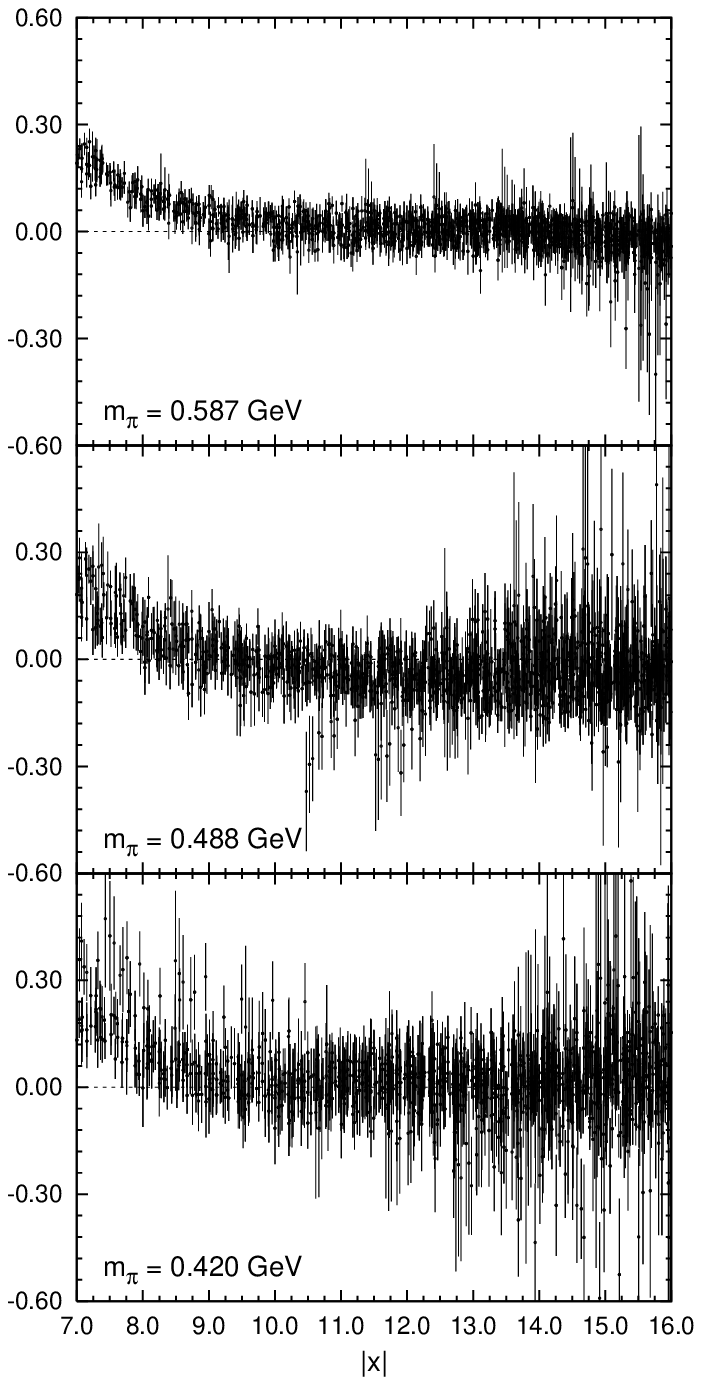}
\end{tabular}
\caption{
$U^{L}_{\rm CM}({\bf x})$ 
as the function of $|{\bf x}|$
for the total momentum 
${\bf P}={\bf 0}$ and 
${\bf P}=(2\pi/L){\bf e}_x$.
}
\label{fig:edp-of-int-range}
\newpage
\end{figure}

We show $U^{L}_{\rm CM}({\bf x})$ 
as the function of $|{\bf x}|$
in Fig.~\ref{fig:edp-of-int-range}.
We find $R< L/2(=16)$ in all cases.
This means that the necessary condition 
for the finite size formula is satisfied 
on the $32^3$ lattice 
for our range of the quark mass $m_\pi = 0.420 - 0.587\ {\rm GeV}$
and the momentum
$k \sim 0 - \pi/(\gamma L) ( \sim 0.16\ {\rm GeV})$
with the current statistics of the simulations.
%
%
\subsection{ Scattering phase shift from wave function } 
\label{sub-sec:Scattering phase from wave function}
We attempt to calculate the scattering phase shift
by substituting 
the momentum $k$ obtained with three methods
into the finite size formula (\ref{eqn:RG-formula}).
First method is the conventional one,
where we calculate the momentum $k$ 
by (\ref{eqn:k2_def})
with the energy of the two-pion system $E$
extracted from the two-pion time correlator.
Our results of $k$ and the scattering phase shift $\delta_0(k)$ 
are tabulated in
Table~\ref{tbl:main-results_0} for ${\bf P}={\bf 0}$ and 
Table~\ref{tbl:main-results_1} for ${\bf P}=(2\pi/L){\bf e}_x$ 
(labeled ``from $T$'').

\begin{figure}[h]
\begin{tabular}{ll}
\qquad\qquad
\qquad
& (a) ${\bf P}={\bf 0}$            \\
& \includegraphics[width=120mm]{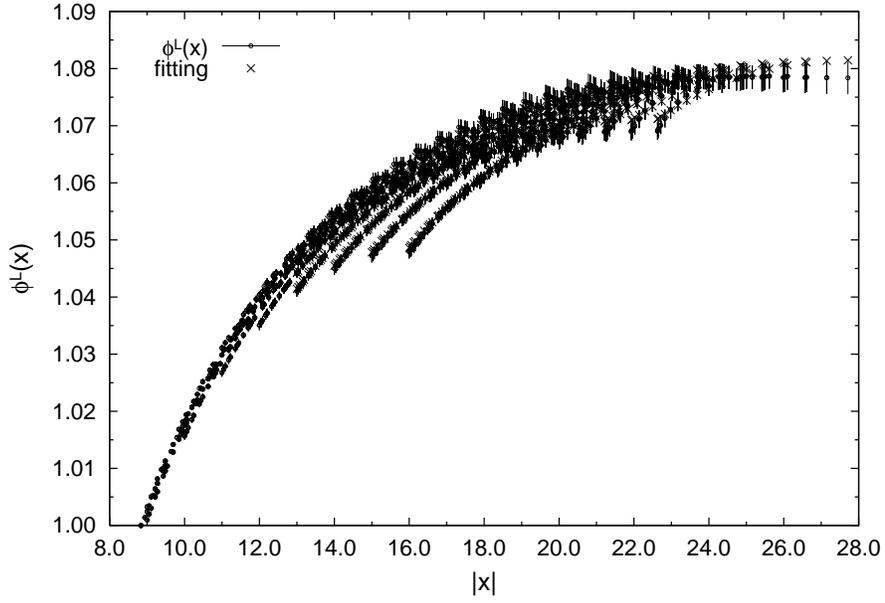} \\ \\
& (b) ${\bf P}=(2\pi/L){\bf e}_x$  \\
& \includegraphics[width=120mm]{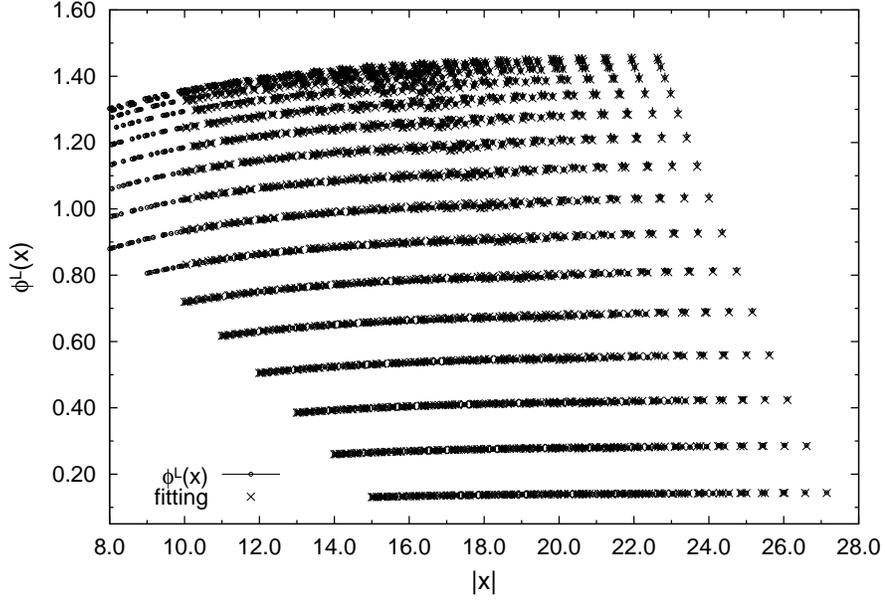}
\end{tabular}
\caption{
Results of fitting
the wave function $\phi^{L}({\bf x})$
to the Green function at $m_\pi=0.420\ {\rm GeV}$
for the total momentum
${\bf P}={\bf 0}$ and 
${\bf P}=(2\pi/L){\bf e}_x$.
Open circles refer to the data points
and cross symbols to the results of the fitting.
}
\label{fig:wave-function-fit}
\newpage
\end{figure}

In the second method,
we extract $k$ by fitting the wave function $\phi^{L}({\bf x})$ 
to the Green function (\ref{eqn:Green_def})
with the form,
\begin{equation}
   \phi^{L}({\bf x}) = C \cdot G(\hat{\gamma}[{\bf x}];k)
\ ,
\label{eqn:W_fit_G_def}
\end{equation}
taking $k$ and an overall constant $C$
as free parameters.
This method was introduced by the CP-PACS collaboration
in Ref.~\cite{Aoki:2005uf}
for the evaluation of the scattering length,
where they found that the statistical error can be reduced
with this method.
The numerical evaluation of $G(\hat{\gamma}[{\bf x}];k)$ 
and an explicit procedure of the fitting
are discussed in
Appendix~\ref{apn:Fitting wave function to Green function}.
We choose the fitting range $|{\bf x}| \ge x_m$
with $x_m$ tabulated in 
Table~\ref{tbl:main-results_0} and \ref{tbl:main-results_1}.
An example of the fitting is shown 
in Fig.~\ref{fig:wave-function-fit}
at $m_\pi = 0.420\ {\rm GeV}$,
where the data points are shown with open circles
and the values for the fits with cross symbols.
We find that the fit works well 
both for ${\bf P}={\bf 0}$ and ${\bf P}=(2\pi/L){\bf e}_x$.
These mean that
the contribution of $G_{lm}({\bf x};k)$ with $l \ge 2$ 
are negligible as expected.
Our results of $k$ and $\delta_0(k)$
are tabulated in 
Table~\ref{tbl:main-results_0} and \ref{tbl:main-results_1} 
(labeled ``from $\phi$'').
These are consistent with
those of the conventional method (``from $T$'').
We also find that 
the statistical errors is reduced.
Thus the method with the wave function 
is also efficient 
for the evaluation of the scattering phase shift.

As discussed in Sec.~\ref{sub-sec:Wave functions},
the ratio $V^{L}_{\rm CM}({\bf x})$ in (\ref{eqn:V_def})
equals to $-k^2$ away from the interaction range.
For simplicity we write the ratio by
\begin{equation}
  V({\bf x};\gamma) \equiv V^{L}_{\rm CM}(\hat{\gamma}[{\bf x}])
      = \frac{1}{ \phi^{L}({\bf x}) }
        \left(
               \frac{1}{\gamma^2} \frac{\partial^2}{\partial x_1^2} 
             +                    \frac{\partial^2}{\partial x_2^2} 
             +                    \frac{\partial^2}{\partial x_3^2} 
        \right) \phi^{L}({\bf x})
\ ,
\label{eqn:V_fit_V_def}
\end{equation}
with (\ref{eqn:V_rewrite}).
We extract $k$ by minimizing the chi-square :
\begin{equation}
   \chi^2(k)
     = \sum_{ |{\bf x}| \ge x_m }
           \left( 
               \frac{        V({\bf x};\gamma) + k^2 }
                    { \delta V({\bf x};\gamma)       }  
           \right)^2
\ ,
\label{eqn:V_fit_const_def}
\end{equation}
taking $k$ as a free parameter,
where $\delta V({\bf x};\gamma)$ 
is the statistical error at fixed $\gamma$.
This method is an extension 
of that introduced by the CP-PACS collaboration
in Ref.~\cite{Aoki:2005uf}
for ${\bf P}={\bf 0}$
to the system with ${\bf P}\ne 0$.
As known in (\ref{eqn:V_fit_V_def}),
$V({\bf x};\gamma)$ depends on $k$ through $\gamma$.
This fact makes the analysis difficult.
We use a similar procedure as for fitting the wave function
(see in Appendix~\ref{apn:Extraction of momentum from V}).
We choose the same fitting range $x_m$ 
for fitting the wave function.
Our results of $k$ and $\delta_0(k)$ 
are tabulated in
Table~\ref{tbl:main-results_0} and \ref{tbl:main-results_1} 
(labeled ``from $V$'').
As shown in the tables,
the results are consistent 
with those from the two-pion time correlator (``from $T$'')
and fitting the wave function (``from $\phi$'').
We also find that
the statistical error is significantly reduced.
%
%
\subsection{ Scattering length and phase shift at physical quark mass } 
\label{sub-sec:Scattering length and phase shift at physical quark mass} 
\begin{figure}[h]
\includegraphics[width=120mm]{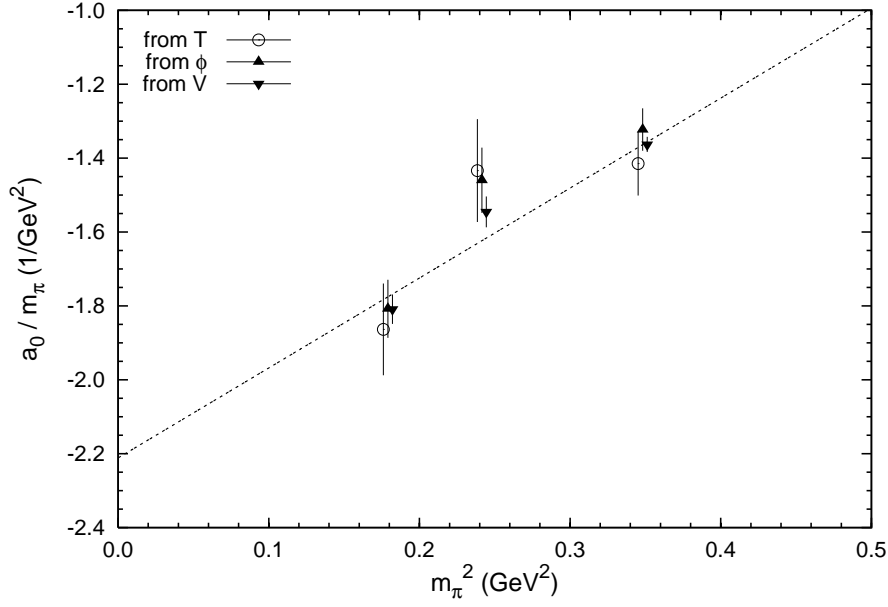}
\caption{
Scattering length $a_0/m_\pi$
obtained from
the two-pion time correlator (``from $T   $''),
fitting the wave function    (``from $\phi$'') and
fitting $V({\bf x};\gamma)$  (``from $V   $'').
The result of chiral extrapolation
for the data of ``from $V$'' is also plotted.
}
\label{fig:mpi2-a0mpi}
\vspace{5mm}
\end{figure}

For ${\bf P}={\bf 0}$,
the momentum $k$ is very small 
as shown in Table~\ref{tbl:main-results_0}.
Thus the scattering length $a_0$ can be calculated by 
\begin{equation}
    a_0 = \frac{ \tan\delta_0(k) }{ k }
\ .
\label{eqn:a0_calc}
\end{equation}
In Fig.~\ref{fig:mpi2-a0mpi}
we plot the scattering length $a_0/m_\pi$ 
obtained from three methods
discussed in the previous section.
They are consistent,
but the statistical errors
are significantly reduced by using the two methods 
with the wave function (``from $\phi$'' and ``from $V$'').
We carry out the chiral extrapolation
with the fit form,
$a_0/m_\pi = A + B \cdot m_\pi^2$
and obtain
\begin{equation}
  a_0 / m_\pi = -2.211(77)\ {\rm GeV}^{-2}
\ ,
\label{eqn:a0_mpi_result}
\end{equation}
in the chiral limit,
for the data of fitting $V({\bf x},\gamma)$ (``from $V$'').
The result of chiral extrapolation
is also plotted in the Fig.~\ref{fig:mpi2-a0mpi}.
We refer that the prediction from CHPT is
$a_0/m_\pi = -2.265(51)\ {\rm GeV}^{-2}$~\cite{CHPT_a0mp}.

\begin{figure}[h]
\includegraphics[width=120mm]{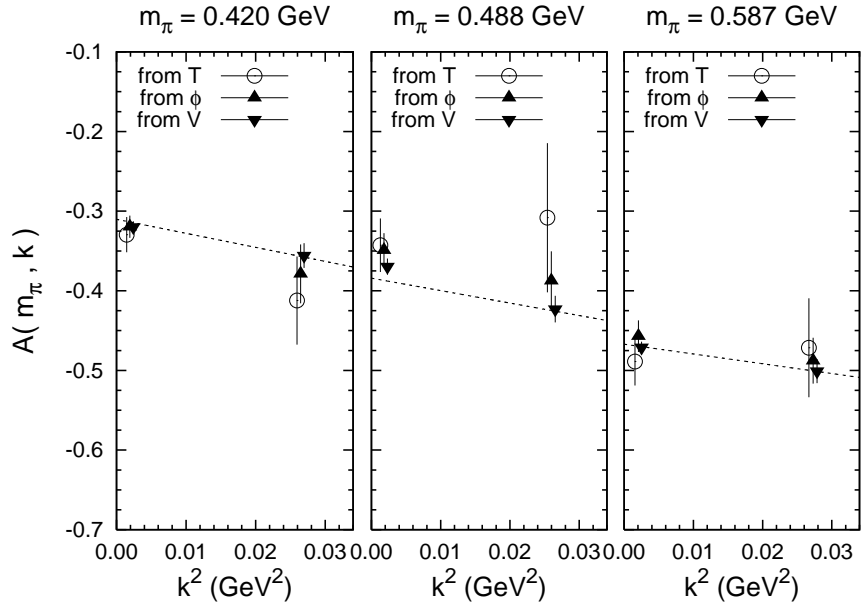}
\caption{
Scattering amplitude $A(m_\pi,k)$ 
fore three quark masses
obtained from
the two-pion time correlator (``from $T   $''),
fitting the wave function    (``from $\phi$'') and
fitting $V({\bf x};\gamma)$  (``from $V   $'').
The fit curves for the data of ``from $V$''
are also plotted.
}
\label{fig:k2-amplitude}
\newpage
\end{figure}

We obtain the scattering phase shift
at the physical quark mass for the various momenta $k$
from the scattering amplitude defined by 
\begin{equation}
  A(m_\pi,k) = \frac{ \tan\delta_0(k) }{ k } 
               \sqrt{ m_\pi^2 + k^2 } 
\ ,
\label{eqn:SCamp_def}
\end{equation}
which is normalized as $A(m_\pi,k)=a_0 m_\pi$ at $k=0$.
We show the results
in Table~\ref{tbl:main-results_0} and \ref{tbl:main-results_1},
and plot in Fig.~\ref{fig:k2-amplitude}.
We fit these amplitudes with a fitting form :
\begin{equation}
  A(m_\pi,k) =  
           A_{10} \cdot m_\pi^2     
         + A_{20} \cdot m_\pi^4
         + A_{01} \cdot k^2
         + A_{11} \cdot m_\pi^2 k^2 
\ .
\label{eqn:SCamp_fit}
\end{equation}
The results of the fitting are also plotted
in Fig.~\ref{fig:k2-amplitude} and 
summarized in Table~\ref{tbl:coefficients}.
We find that $A_{10}(=a_0/m_\pi)$ is consistent 
with (\ref{eqn:a0_mpi_result}).

\begin{figure}[t]
\includegraphics[width=120mm]{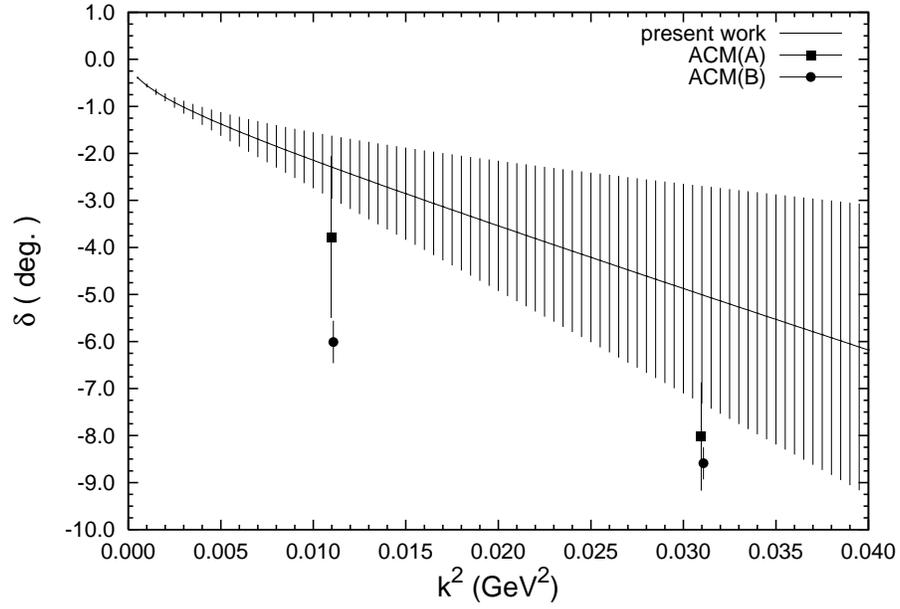}
\caption{
Results of the scattering phase shift
at the physical pion mass
and the experiments in Ref.~\cite{Hoogland:1977kt}.
}
\label{fig:delta-at-physical-pion}
\newpage
\end{figure}

Finally,
we estimate the scattering phase shift $\delta_0(k)$ 
at the physical quark mass by putting $m_\pi=0.140$ GeV
in (\ref{eqn:SCamp_fit}).
In Fig.~\ref{fig:delta-at-physical-pion}
our results are plotted as a function of $k^2$
and compared with the experiments~\cite{Hoogland:1977kt}.
Our results are larger than the experiments.
We consider that this disagreement
arises from the effect of the finite lattice spacing.
We must leave this issue to studies in the future.
%
%
\section{Conclusion}
\label{sec:conclusion}
We have calculated the two-pion wave function 
for the ground state of the $I=2$ $S$-wave two-pion system 
both for ${\bf P}={\bf 0}$ and ${\bf P}=(2\pi/L){\bf e}_x$.
We have investigated the validity of 
the necessary condition for the finite size formula 
and found that it is satisfied 
on the $32^3$ lattice 
for the quark mass range $m_\pi = 0.420 - 0.587 \ {\rm GeV}$ 
and the scattering momentum $k \le 0.16 \ {\rm GeV}$. 
We have also found 
that the scattering phase shift 
can be extracted from the 
wave function with a smaller statistical error
than from the two-pion time correlator,
which have been used in the studies to date.

An implication of the present work
is the feasibility to calculate the decay width of the $\rho$ meson
through studies of the $I=1$ two-pion system.
While the evaluation of the disconnected diagrams
with a good precision has been a computational problem,
our method, investigating the scattering system
from the wave function
in which the energy eigenvalue is extracted
from the wave function at a single time slice,
could lend a tactics that
can be used to evaluate such complicated diagrams
with a modest cost.
%
%
\section*{ Acknowledgments }
We acknowledge to
Y.Namekawa for helping us to develop a solver program. 
The numerical calculations have been performed on VPP5000/80 at the
Academic Computing and Communications Center of University of Tsukuba.
%
%
\clearpage
\appendix
\section{ Fitting wave function to Green function }
\label{apn:Fitting wave function to Green function}
We introduce
\begin{equation}
   \bar{k}^2 = ( \hat{\gamma}^{-1}[{\bf P}/2] )^2
\ , 
\end{equation}
and expand the Green function around
$k^2 = \bar{k}^2$ as
\begin{equation}
  G(\hat{\gamma}[{\bf x}];k)
     = \sum_{j=0}^{\infty}
         ( k^2 - \bar{k}^2 )^{j-1}
         \cdot F(\hat{\gamma}[{\bf x}];j,\bar{k})
\ .
\label{eqn:Green_exp_F}
\end{equation}
The coefficient $F({\bf x};j,\bar{k})$ is defined by 
\begin{equation}
  F({\bf x};0,\bar{k})
  = - \frac{1}{\gamma L^3}
      \sum_{ {\bf q}^2 = \bar{k}^2 }
              {\rm e}^{ i{\bf q}\cdot{\bf x} }
\ ,
\label{eqn:Green_exp_F0_def}
\end{equation}
and 
\begin{equation}
  F({\bf x};j,\bar{k})
  = \frac{1}{\gamma L^3}
       \sum_{ {\bf q}^2 \ne \bar{k}^2 } 
          \frac{1}{ ( {\bf q}^2 - \bar{k}^2 )^{j} }
          {\rm e}^{ i{\bf q}\cdot{\bf x} }
\ ,
\label{eqn:Green_exp_F1_def}
\end{equation}
for $j\ge 1$,
where ${\bf q}\in \Gamma_{\bf P}$
(defined by (\ref{eqn:def_Gamma})).
We evaluate $F({\bf x};j,\bar{k})$ for $j\ge 1$
from
\begin{eqnarray}
\lefteqn{ (\gamma L^3) \cdot F({\bf x};j,\bar{k}) } \cr
&=&
  -
         \frac{1}{j!} 
         \sum_{ {\bf q}^2 = \bar{k}^2 }
                {\rm e}^{ i {\bf q}\cdot{\bf x} }
  +
         \sum_{r=1}^j 
         \frac{1}{(j-r)!}
         \sum_{ {\bf q}^2 \ne \bar{k}^2 }
             \frac{ {\rm e}^{ - ( {\bf q}^2 - \bar{k}^2 ) } }
                  { ( {\bf q}^2 - \bar{k}^2 )^r }
         {\rm e}^{i {\bf q}\cdot{\bf x}}
\cr
& &
  +
    \frac{ 2 \pi^{3/2} \gamma }{ (j-1)! }
    \left( \frac{2\pi}{L} \right)^{-3}
    \int_{0}^{1} {\rm d}\eta\ 
           \eta^{2j-4} {\rm e}^{ (\eta \bar{k})^2 }
           \cdot f( {\bf x};\eta )
\ ,
\label{eqn:Green_exp_F1_calc}
\end{eqnarray}
where
\begin{equation}
    f( {\bf x}; \eta )
       = 
       \sum_{ {\bf n}\in\mathbb{Z}^3 }
          (-1)^{ L/(2\pi) {\bf P}\cdot{\bf n} }
          \cdot
          {\rm e}^{ - \bigl( ( {\bf x} + \hat{\gamma}[ {\bf n} ] L ) / (2\eta) \bigr)^2}
\ .
\label{eqn:Green_exp_F1_calc_sub}
\end{equation}
(\ref{eqn:Green_exp_F1_calc}) 
is obtained by the same technique discussed 
in Ref.~\cite{Aoki:2005uf}.
$F(\hat{\gamma}[{\bf x}];j,\bar{k})$ in the (\ref{eqn:Green_exp_F})
is depend on $k$ through $\gamma$.
This fact makes the fitting of the wave function difficult.
In the present work,
we adopt the following procedure:  
%
\renewcommand{\labelenumi}{[\arabic{enumi}]} 
\begin{enumerate} 
%
\item \label{proc_W0}
We calculate
$\gamma_0 = E/\sqrt{E^2-{\bf P}^2}$
with the energy $E$
extracted from the two-pion time correlator
and take it as the initial value of $\gamma$.
%
\item \label{proc_W1}
We fix $\bar{k}$ and 
$F(\hat{\gamma}[{\bf x}];j,\bar{k})$ in the (\ref{eqn:Green_exp_F})
with given $\gamma$ and
fit the wave function,
taking $k$ as the free parameter.
%
\item \label{proc_W2}
We update $\gamma$ by 
$\gamma = \sqrt{ 1 + {\bf P}^2/[4(m_\pi^2+k^2)] }$
with $k$ obtained in [\ref{proc_W1}].
%
\item \label{proc_W3}
We iterate the procedure [\ref{proc_W1}] - [\ref{proc_W2}],
until the value of $k$ becomes to be stable.
\end{enumerate} 
%
It is expected that $k$ rapidly converges in this procedure,
because the dependence of $\gamma$ on $k$ is very small.
We confirm that $k$ is stable 
within the single precision after five iterations
for all our simulation parameters.
We carry out above procedure for the each jackknife bin
and estimate the statistical error of $k$ 
by the jackknife method.
%
%
\section{ Fitting $V({\bf x};\gamma)$ }
\label{apn:Extraction of momentum from V}
The fit of $V({\bf x};\gamma)$
is performed by the similar procedure 
as for fitting the wave function 
discussed in
Appendix~\ref{apn:Fitting wave function to Green function} :
%
\renewcommand{\labelenumi}{[\arabic{enumi}]} 
\begin{enumerate} 
%
\item \label{proc_V0}
We calculate
$\gamma_0 = E/\sqrt{E^2-{\bf P}^2}$
with the energy $E$
extracted from the two-pion time correlator
and take it as the initial value of $\gamma$.
%
\item \label{proc_V1}
We fix 
$V({\bf x};\gamma)$ with given $\gamma$ 
and 
search $k$ that gives the least number of $\chi^2(k)$ 
in (\ref{eqn:V_fit_const_def}).
%
\item \label{proc_V2}
We update $\gamma$ by 
$\gamma = \sqrt{ 1 + {\bf P}^2/[4(m_\pi^2+k^2)] }$ 
with $k$ obtained in [\ref{proc_V1}].
%
\item \label{proc_V3}
We iterate the procedure [\ref{proc_V1}] - [\ref{proc_V2}],
until the value of $k$ becomes to be stable.
\end{enumerate}
%
It is expected that $k$ in this procedure rapidly converges,
because the dependence of $\gamma$ on $k$ is very small.
We confirm that $k$ is stable 
within the single precision after five iterations
as for the fitting the wave function.
We carry out above procedure for the each jackknife bin
and estimate the statistical error of $k$ by the jackknife method.
%
%

\newpage
%
%
%
%
%
%
%
%
%
%
%
%
%
%
%
%
\begin{table}[h]
\begin{ruledtabular}
\begin{tabular}{ c c c c c }
$ \kappa $                &  $ N_{\rm conf.} $         &  
$ m_\pi \ ({\rm GeV}) $   &  $ m_\rho\ ({\rm GeV}) $   &  $ m_\pi / m_\rho $  \\
\hline
$ 0.1413      $   &   $ 400        $   &
$ 0.41964(66) $   &   $ 0.837(12)  $   &   $ 0.5012(71) $  \\
$ 0.1410      $   &   $ 212        $   &
$ 0.48829(95) $   &   $ 0.8598(95) $   &   $ 0.5679(62) $  \\
$ 0.1405      $   &   $ 212        $   &  
$ 0.58749(74) $   &   $ 0.9007(67) $   &   $ 0.6522(48) $  \\
\end{tabular}
\end{ruledtabular}
\caption{
Masses of the pion and the $\rho$ meson
calculated from the time correlator.
}
\label{tbl:mass-param}
\vspace{10mm}
\end{table}
%
%
\begin{table}[h]
\begin{ruledtabular}
\begin{tabular}{lllll}
$ \kappa $
&              &  $  0.1413      $  &  $ 0.1410       $  &  $ 0.1405       $  \\
$ m_\pi\ ({\rm GeV}) $
&              &  $ 0.41964(66)  $  &  $ 0.48829(95)  $  &  $ 0.58749(74)  $  \\
\hline
$ x_m $
&              &  $ 13.0         $  &  $ 14.0         $  &  $ 12.0         $  \\
$ k \ ({\rm GeV}) $  \\
& from $T   $  &  $ 0.0383(14)   $  &  $ 0.0360(19)   $  &  $ 0.0396(14)   $  \\
& from $\phi$  &  $ 0.03760(91)  $  &  $ 0.0363(12)   $  &  $ 0.03810(92)  $  \\
& from $V   $  &  $ 0.03762(46)  $  &  $ 0.03750(57)  $  &  $ 0.03875(33)  $  \\
$\delta_0(k) \ ({\rm deg.}) $  \\
& from $T   $  &  $ -1.71(18)    $  &  $ -1.44(22)    $  &  $ -1.88(18)    $  \\
& from $\phi$  &  $ -1.63(11)    $  &  $ -1.48(14)    $  &  $ -1.70(11)    $  \\
& from $V   $  &  $ -1.636(56)   $  &  $ -1.621(69)   $  &  $ -1.776(42)   $  \\
$ a_0/m_\pi \ ({\rm GeV}^{-2}) $  \\
& from $T   $  &  $ -1.86(12)    $  &  $ -1.43(14)    $  &  $ -1.415(87)   $  \\
& from $\phi$  &  $ -1.808(79)   $  &  $ -1.460(88)   $  &  $ -1.322(58)   $  \\
& from $V   $  &  $ -1.809(40)   $  &  $ -1.546(42)   $  &  $ -1.363(21)   $  \\
$ A(m_\pi,k) $  \\
& from $T   $  &  $ -0.330(22)   $  &  $ -0.343(34)   $  &  $ -0.489(30)   $  \\
& from $\phi$  &  $ -0.320(14)   $  &  $ -0.349(21)   $  &  $ -0.457(20)   $  \\
& from $V   $  &  $ -0.3198(70)  $  &  $ -0.370(10)   $  &  $ -0.4709(72)  $
\end{tabular}
\end{ruledtabular}
\caption{
Results for the total momentum ${\bf P}={\bf 0}$
from two-pion time correlator (``from $T   $''),
fitting wave function         (``from $\phi$'') and 
fitting $V({\bf x};\gamma)$   (``from $V   $'').
}
\label{tbl:main-results_0}
\newpage
\end{table}
%
%
\begin{table}[h]
\begin{ruledtabular}
\begin{tabular}{lllll}
$ \kappa $
&              &  $ 0.1413       $  &  $ 0.1410       $  &   $ 0.140        $  \\
$ m_\pi\ ({\rm GeV}) $
&              &  $ 0.41964(66)  $  &  $ 0.48829(95)  $  &   $ 0.58749(74)  $  \\
\hline
$ x_m $
&              &  $ 10.0         $  &  $ 11.0         $   &  $ 12.0         $  \\
$ k \ ({\rm GeV}) $  \\
& from $T   $  &  $ 0.1610(14)   $  &  $ 0.1594(21)   $   &  $ 0.1634(12)   $  \\
& from $\phi$  &  $ 0.16015(96)  $  &  $ 0.16117(82)  $   &  $ 0.16369(54)  $  \\
& from $V   $  &  $ 0.15956(40)  $  &  $ 0.16197(38)  $   &  $ 0.16394(28)  $  \\
$ \gamma $ \\
& from $T   $  &  $ 1.06159(22)  $  &  $ 1.04749(23)  $   &  $ 1.033921(92) $   \\
& from $\phi$  &  $ 1.06168(19)  $  &  $ 1.04739(17)  $   &  $ 1.033944(82) $   \\
& from $V   $  &  $ 1.06173(17)  $  &  $ 1.04734(16)  $   &  $ 1.033937(82) $   \\
$\delta_0(k) \ ({\rm deg.}) $  \\
& from $T   $  &  $ -8.4(12)     $  &  $ -5.5(17)     $   &  $ -7.20(98)    $  \\
& from $\phi$  &  $ -7.69(78)    $  &  $ -6.92(69)    $   &  $ -7.46(46)    $  \\
& from $V   $  &  $ -7.21(33)    $  &  $ -7.59(31)    $   &  $ -7.67(24)    $  \\
$ A(m_\pi,k) $  \\
& from $T   $  &  $ -0.412(55)   $  &  $ -0.308(94)   $   &  $ -0.472(62)   $  \\
& from $\phi$  &  $ -0.379(37)   $  &  $ -0.387(37)   $   &  $ -0.488(29)   $  \\
& from $V   $  &  $ -0.356(15)   $  &  $ -0.423(17)   $   &  $ -0.501(15)   $
\end{tabular}
\end{ruledtabular}
\caption{
Results for the total momentum ${\bf P}=(2\pi/L){\bf e}_x$
from two-pion time correlator (``from $T   $''),
fitting wave function         (``from $\phi$'') and 
fitting $V({\bf x};\gamma)$   (``from $V   $'').
}
\label{tbl:main-results_1}
\vspace{10mm}
\end{table}
%
%
\begin{table}[h]
\begin{ruledtabular}
\begin{tabular}{lllll}
&  \multicolumn{1}{c}{$ A_{10} \ ({\rm GeV}^{-2})$}
&  \multicolumn{1}{c}{$ A_{20} \ ({\rm GeV}^{-4})$}
&  \multicolumn{1}{c}{$ A_{01} \ ({\rm GeV}^{-2})$}
&  \multicolumn{1}{c}{$ A_{11} \ ({\rm GeV}^{-4})$} \\
\hline
from $T   $  &  $ -2.10(27)  $  &  $ 2.02(92) $  &  $ -7.7(56) $  &  $ 26(22)  $  \\
from $\phi$  &  $ -2.17(17)  $  &  $ 2.52(59) $  &  $ -3.2(34) $  &  $ 6(12)   $  \\
from $V   $  &  $ -2.187(80) $  &  $ 2.42(26) $  &  $ -2.3(15) $  &  $ 3.2(56) $  \\
\end{tabular}
\end{ruledtabular}
\caption{
Results of the coefficients $A_{ij}$ 
in (\ref{eqn:SCamp_fit}).
}
\label{tbl:coefficients}
\end{table}
%
%
\end{document}